\documentclass[manuscript]{aastex}
\usepackage{graphicx}
\usepackage{natbib}
\usepackage{rotating}
\slugcomment{To be submitted to the Astrophysical Journal soon...}

\shorttitle{The Asymmetric Wind of WR1}
\shortauthors{St-Louis}

\begin{document}

\title{Revealing the Asymmetry of the Wind of the Variable Wolf-Rayet Star WR1 (HD4004) Through Spectropolarization}

\author{N. St-Louis}
\affil{D\'epartement de physique and Centre de Recherche en Astrophysique du Qu\'ebec (CRAQ), Universit\'e de Montr\'eal, C.P. 6128, Succ. Centre Ville, Montr\'eal (Qc), H3C 3J7, Canada: stlouis@astro.umontreal.ca}

\begin{abstract}
In this paper, high quality spectropolarimetric observations of the Wolf-Rayet (WR) Star WR1 (HD4004) obtained with ESPaDOnS 
at CFHT are presented.  All major emission lines present in the spectrum show depolarization in the relative Stokes
parameters $Q/I$ and $U/I$. From the behaviour of the amount of line depolarization as a function of line strength, the intrinsic continuum light polarization of WR1 is estimated to be $P/I=0.443\pm0.028\%\ $ with an angle of $\theta=-26.2^{\circ}$.

Although such a level of polarization could in principle be caused by a wind flattened by fast rotation, the scenario in which it is a 
consequence of the presence in the wind of Corotating Interaction Regions (CIRs) is preferred. This is supported by previous 
photometric and spectroscopic observations showing periodic variations with a period of 16.9 days. This is now the third
WR star thought to exhibit CIRs in its wind that is found to have line depolarization. Previous authors have found a strong correlation between line depolarization and the presence of an ejected nebula, which they interpret as a sign that the star has reached the WR phase relatively recently since the nebula are thought to dissipate very fast. In the cases where the presence of CIRs in the wind is  favoured to explain the depolarization across spectral lines, the above mentioned correlation may be indicating that those massive stars have only very recently transited from the previous evolutionary phase to the WR phase.

\end{abstract}


\section{The Structure of Massive-Star Winds}

The phenomenal mass-loss from massive stars from their birth to their death is 
a crucial element of our understanding of the ecology of galaxies. The incessant exchange 
of gas between stars and the interstellar medium (ISM) is a determining factor of galaxy evolution,
and massive stars play a particularly important role with regards to energy and momentum 
input, as well as in the production of chemically processed material. The amount of mass
lost in each evolutionary phase but also the way it is lost, i.e. the geometry of the wind, will 
determine to what extent the ISM will be affected and in what way.

Equally important is what can be learned about massive stars themselves by studying 
their dense winds. Over the past decades, theoretical investigations on the effect
of mass-loss \citep[e.g.][]{1981A&A...102..401M,1986ARA&A..24..329C,1993ApJ...411..823W}, metallicity 
\citep[e.g.][]{1994A&AS..103...97M}, rotation \citep[e.g.][]{2000ARA&A..38..143M,2011A&A...530A.115B} and magnetism 
\citep[e.g.][]{2004A&A...422..225M, 2011A&A...525L..11M,2012A&A...542A.113Y} on the evolution 
of massive stars have been carried out \citep[see a recent review on massive-star 
evolution by][]{2012ARA&A..50..107L}. Observationaly, we have only very limited information
on the two last parameters and still have many things to learn about the first two. As the wind is 
initiated near the stellar photosphere, the outflow will ultimately be affected by physical processes 
originating at or close to the surface of the star. Therefore, studying the stellar mass-loss has the 
potential of helping us secure new information on physical processes for which we know very 
little at the present time and which have a significant impact on the life of the star. 
The geometry of the wind is as important a parameter as the amount of material lost.
\citet{1995ApJ...440..308C} and \citet{1999A&A...347..185M} found that when gravity darkening is included, the wind driving is higher at the poles than at the equator. This will inevitably lead to a reduced loss of angular momentum and therefore of the star's rotation rate compared to the case where the wind is spherically symmetric. Nevertheless, \citet{2000A&A...361..101M} still predict a large decrease of surface velocities on the main sequence and an even larger one when the star reaches the Wolf-Rayet (WR) phase for rotating massive stars. Any other type of wind asymmetry would have to be modelled in detail to assess how it affects the loss of angular momentum.

For WR stars, the situation is particularly unclear. At the present time, only upper limits have been set for 
magnetic fields in WR stars \citep{2013ApJ...764..171D}. The difficulty arises because there are no photospheric lines
in the spectrum of these stars and therefore we are forced to study emission lines formed in the thick wind \citep[see predicted Stokes $V$ profiles for emission lines formed in hot-star winds threaded with a weak split monopole magnetic field in][]{2010ApJ...708..615G}.  Constraints on magnetic fields in the wind can in principle be extrapolated back to the stellar surface, 
but this is very model dependent. The lack of phostopheric lines also prevents the direct determination 
of the rotation rate of the star by the usual method of measuring the line broadening. Indirect means 
have been used in the case of the presence of periodic variations in presumably single stars \citep[e.g.][]{2010ApJ...716..929C} but some uncertainties still remain.

The above-mentioned physical processes, combined with the basic radiative driving of the wind, determine its global density distribution. Initially, for simplicity, the winds were assumed to be smooth and have spherical symmetry. Not surprisingly, 
several observational discoveries were to prove this wrong. Photometric and spectral variations, not readily 
associated with easily identifiable massive WR+O binaries, were found. To explain these, suggestions such 
as variability associated with the presence of a compact companion \citep{1982IAUS...99..263M, 1983wrsp.conf...13M} or non-radial pulsations \citep{1985PASP...97..274V} were put forward but none were found to be completely satisfactory, mainly because the changes were not found to be {\it strictly} periodic.

Later, small-scale emission excesses were found superposed on the broad emission lines formed in the wind of several Wolf-Rayet stars \citep{1988ApJ...334.1038M, 1999ApJ...514..909L}.  These were eventually associated with clumps in the wind 
moving out radially. It is now thought that clumping is a universal parameter of massive-star winds \citep[e.g.][]{2008AJ....136..548L}.
These clumps cause non-periodic photometric, polarimetric and spectroscopic variability at a relatively low 
level. They have led to a downward revision of the mass-loss rates of O and WR stars by a factor of 2-5 \citep{1998A&A...333..956N, 1998A&A...335.1003H, 2005A&A...438..301B, 1988ApJ...330..286S}, as the main observational diagnostics depend on the density squared.

With the launch of the {\it International Ultraviolet Explorer} (IUE), another observational feature that was 
found to be ubiquitus was the presence of Discrete Absorption Components (DACs) in the P~Cygni 
profiles of O stars \citep{1989ApJS...69..527H}. \citet{1986A&A...165..157M} was the first to point out that these features can be understood as a consequence of the presence in O-star winds of large-scale structures called Corotating Interaction Regions (CIRs), as first observed in the Sun. These spiral-shaped 
features are formed when a perturbation at the base of the wind, at the hydrostatic radius, causes 
extra radiation pressure which generates high density, low velocity regions which are then carried 
around by the rotation of the star \citep[see the hydrodynamical simulations of][]{1996ApJ...462..469C}. The DACs are formed when these high density zones are seen 
in projection in front of the star and appear as a narrow absorption feature in the major ultraviolet
 P~Cygni absorption troughs and are found to gradually move towards higher velocities and narrower widths. CIRs are also an excellent candidate to explain optical spectroscopic variability observed in the strong emission lines of several presumably single WR stars \citep{2002A&A...395..209D}. In that case, the changes related to CIRs can be periodic but depending on the lifetime of a given CIR and the number of CIRs present,  the phasing or even the general variability pattern can be different at different epochs. 

Linear spectropolarimetry of WR stars is a powerful tool to reveal global wind asymmetries. Indeed, 
linear polarization in these wind is readily generated from Thomson scattering of continuum light, formed 
near the base of the wind, off free electrons in this very hot and highly ionized environment. In a spherically 
symmetric and unresolved outflow, all polarization vectors exactly cancel out, but in the presence of a global asymmetry, 
a net continuum polarization remains. The line flux will be less polarized, if at all. Indeed, most optical
emission lines are formed by recombination which produces unpolarized light. For lines that have a scattering 
component, some polarization can be present, but the net value will always be smaller than that of the 
continuum because the lines are formed further out in the wind where the density of free electrons is smaller. 
The further out they form, the less they are expected to be polarized. Consequently, in asymmetric winds, 
the relative Stokes parameters are expected to show {\it depolarization} across spectral lines.  

In this paper, spectropolarimetric observations of the WN4b star WR~1= HD4004 are presented. This star was 
identified as a good candidate for harbouring CIRs in its wind by \citet{2009ApJ...698.1951S} based on the large-amplitude 
spectroscopic variability it presented. Follow-up observations by \citet{2010ApJ...716..929C} showed that the changes 
were indeed periodic in spectroscopy and photometry with a period of P=16.9 days. The kinematical characteristics
of the spectroscopic variations led to the conclusion that they were generated by the presence of CIRs in the wind. 
Therefore, this  was an ideal candidate to look for wind asymmetries via linear spectropolarization.  In Section 2, the 
observations and data reduction procedures are presented and results are discussed in Section 3. 
The conclusions are given in Section 4.

\section{ESPaDOnS Observations and Data Reduction}

The observations in Stokes $Q$, $U$ and $I$ of WR~1 were obtained  on September 28, 2009 at the 
Canada-France-Hawaii Telescope (CFHT) using ESPaDOnS\footnote{www.cfht.hawaii.edu/Instruments/Spectroscopy/Espadons/},
\emph{Echelle Spectro-Polarimetric Device for the Observations of Stars}, which consists of 
a cross-dispersed, bench-mounted echelle spectrograph combined with a polarimeter. The polarization unit 
is composed of one fixed quarter- and two rotatable half-wave Fresnel rhombs with a Wollaston prism, which 
provides achromatic polarization.  This high resolution (R$\sim$65000), fibre-fed spectropolarimeter is designed 
to obtain a complete optical spectrum (from 3700~\AA\  to 10500~\AA) 
in a single exposure.

Both the $Q$ and $U$ observations discussed in this work consist of four 650~s successive exposures in 
different configurations of the retarder. For the second and third position, the optics settings are changed 
so as to exchange the positions of the two spectra on the CCD.  This allows a minimization of the systematic 
errors due to misalignments, transmission, seeing effects, etc.  From the two sets of 4 sub-exposures, the 
relative polarization, $Q/I$ and $U/I$, are derived using the double-ratio method described by \citet{1997MNRAS.291..658D}. 

The data reduction was carried out by CFHT staff using the Upena pipeline, which is based on Jean-Fran\c{c}ois 
Donati's reduction package Libre-ESpRIT. Libre-ESpRIT, a self-contained package developed specifically 
for reducing echelle spectropolarimetric data, was first introduced by \citet{1997MNRAS.291..658D} and later upgraded
in order to be used specifically with ESPaDOnS.  In this particular case, both the automatic spectrum 
rectification and the continuum polarization removal options have been disabled. This is because the 
code was not designed to work with spectra containing many strong and broad emission lines from hot star winds 
and will therefore not perform the data reduction correctly if these options are used.

The reduction of ESPaDOnS data\footnote{http://www.cfht.hawaii.edu/Instruments/Spectroscopy/Espadons/Espadons\_esprit.html}  
begins by a geometrical analysis of each order from calibration exposures. The wavelength-to-pixel solution 
is obtained from a comparison frame (Thorium). From this,  an optimal extraction of the spectrum is obtained. 
In this process, two check spectra ($N_1$ and $N_2$) and error bars at each wavelength point in the spectrum are also produced. 
The former are \emph{null polarization} profiles used to help identify spurious polarization signatures. 
In the final spectrum, each data point corresponds to a velocity bin of 1.8 km~s$^{-1}$.

Before the linear polarization analysis can be performed, a series of operations have been applied to 
optimize the ESPaDOnS data. First, adjacent orders were merged to produce one continuous spectrum 
based on the uncertainty profile supplied by Libre-Esprit. The method used is based on that introduced 
by \citet{1996SoPh..164..417S} in which the order junctions are set at wavelengths where the uncertainty profiles 
from two successive orders cross each other. Note that Libre-Esprit always supplies the absolute error bars even 
though in our case (data reduction without continuum polarization removal) it is the {\it relative} Stokes parameters 
that are provided. Since the polarization level is very small, the error bar on the relative Stokes parameters is simply obtained by dividing the
error bar on the absolute Stokes parameters $Q$ and $U$  by the total intensity $I$ (the exact value from error propagation should be: 
${\sigma \over I}\sqrt{1 + {Q^2 \over I^2}}$ and ${\sigma \over I}\sqrt{1 + {U^2 \over I^2}}$ for $Q$ and $U$ respectively but the values of $Q$ and $U$ are very small compared to $I$). 

The typical noise level from the raw counts of an individual velocity bin is of the order of 10\%. As the typical 
line depolarization values that have been found in the past for WR stars are of the order of 0.5\%, it is impossible 
to distinguish a signal visually. Therefore data binning has been performed in order to increase the S/N of the 
data. As WR1 has several emission lines of different intensities in its spectrum, the signal-to-noise ratio per pixel 
is not constant. In order to obtain similar error bars along the spectrum, a variable binning procedure was applied 
as described in \citet{2013ApJ...764..171D}. Each final bin does not contain the same number of original bins but 
they all have the same error bar. No information is lost in this process if the bins are chosen to be sufficiently narrow. Indeed, 
for WR stars, no feature is expected to be narrower than the typical turbulence in the wind which is estimated at around 
100 km~s$^{-1}$  \citep{1999ApJ...514..909L}.

\section{Spectral Line Depolarizarion in WR1}

In Figures~1 and 2, the Stokes parameters relative to the total flux (line $+$ continuum), $Q/I$ and $U/I$, between 4400 
and 7300 \AA\  are presented in the middle panels while the total flux rectified to the continuum, $I/I_c$, 
is shown in the bottom panels. The null spectra $N_1/I$ are presented in the top panels. The intensity spectrum presented 
here was generated by combining the 8 sub-exposures 
(4 for $Q$ and 4 for $U$) and then rectified using carefully selected continuum regions spread over the entire wavelength 
interval (4420-4440 \AA; 4770-4800 \AA; 4901-4910 \AA; 4975-5130 \AA;  5540-5730 \AA; 5940-6080 \AA; 6735-6830 \AA).  For 
shorter wavelengths, the flux level was too low to obtain small enough bin sizes to distinguish a polarization signal within 
spectral lines, while for longer wavelengths, the large number  of telluric lines, often superposed on stellar emission lines, 
greatly reduces the usefulness of the polarimetric data. Here the $y$-axis for the rectified flux is truncated at 4 continuum units for clarity but 
the He{\sc ii}$\lambda$4686 line reaches 10 continuum units.

\begin{sidewaysfigure}[!Hhtb]
\includegraphics[angle=90,scale=0.75]{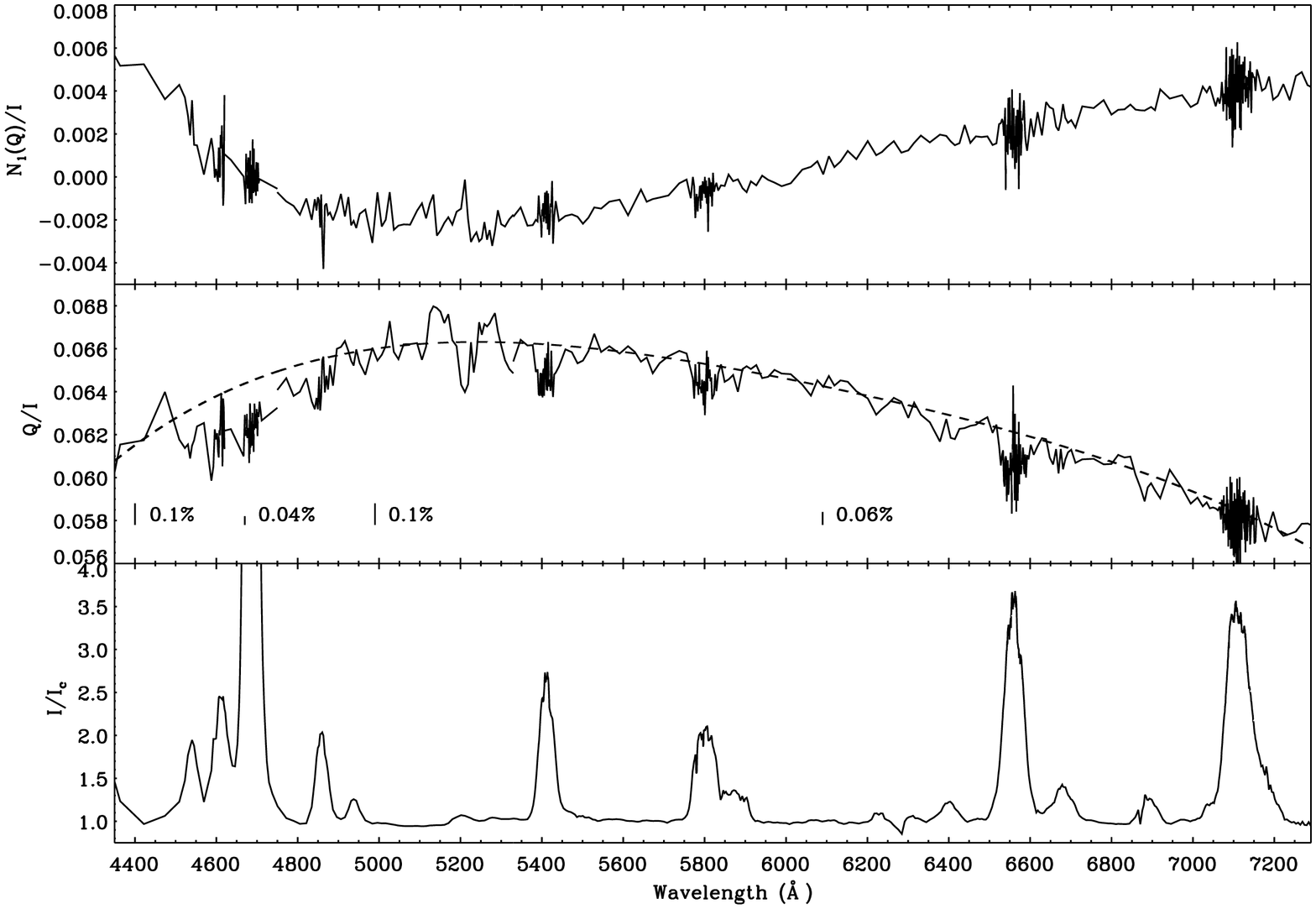}
\caption{Relative Stokes parameter $I/I_c$ (bottom), $Q/I$ (middle) and check spectra $N_1(Q)/I$ (top) for the WR star WR~1. Binned 2$\sigma$ error bars are variable along the spectra and are indicated
in the middle panel.}
\end{sidewaysfigure}

\begin{sidewaysfigure}[!Hhtb]
\includegraphics[angle=90,scale=0.75]{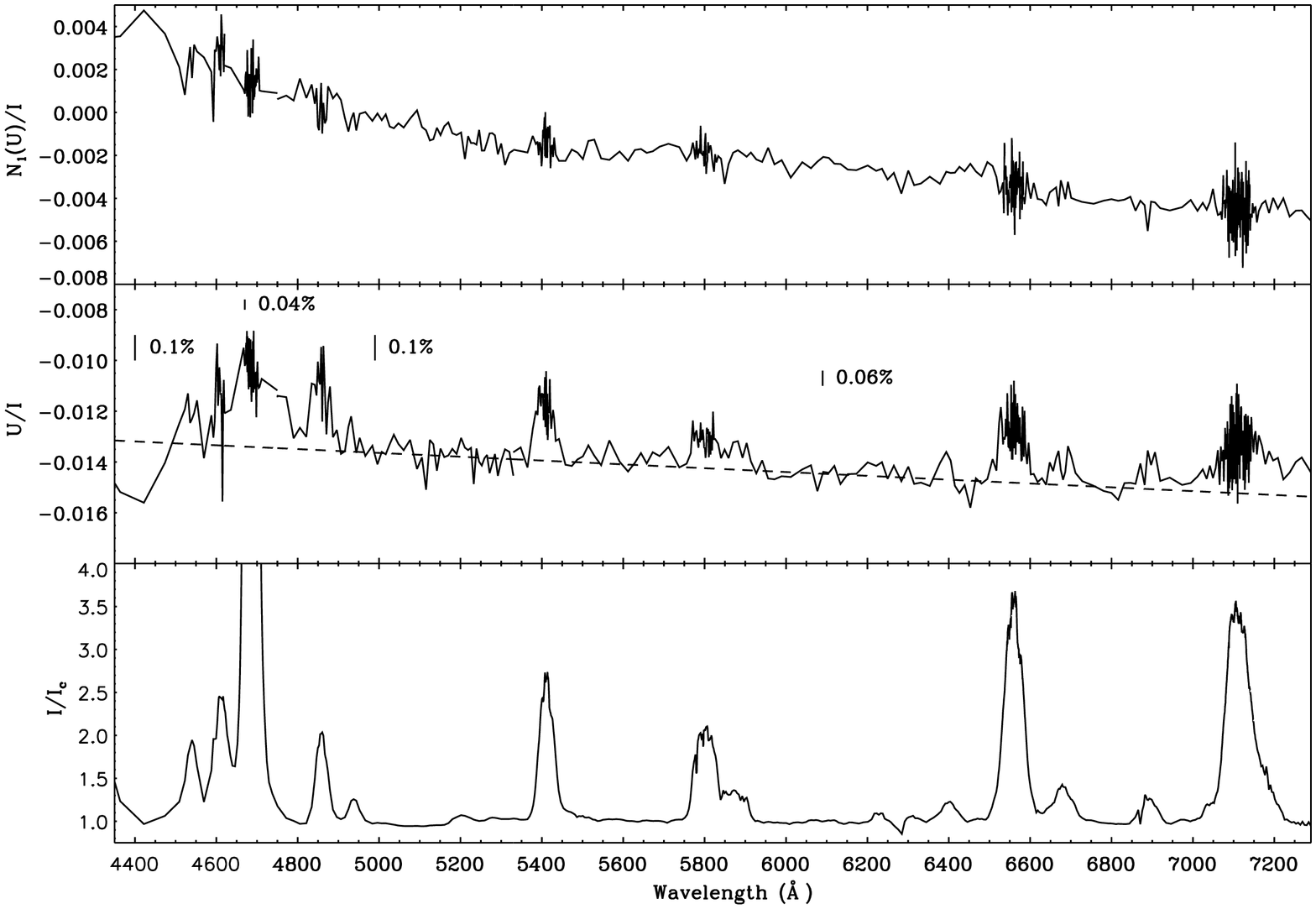}
\caption{Relative Stokes parameter $I/I_c$ (bottom), $U/I$ (middle) and check spectra $N_1(U)/I$ (top) for the WR star WR~1. Binned 2$\sigma$ error bars are variable along the spectra and are indicated
in the middle panel.}
\end{sidewaysfigure}

The signal-to-noise of the data is variable as a function of wavelength. Therefore, in this plot, a different binning of the linear 
polarization flux was adopted for different parts of the spectrum. For data above $\sim$5330 \AA, an error bar of 0.03\%\ is 
used while at shorter wavelengths, it is 0.05\%, except for the He{\sc ii}$\lambda$4686 line where a value of 0.02\%\ was adopted.
Two sigma error bars are drawn in the middle panels ($Q/I$ and $U/I$ plots). 
Of course, the nature of the binning process is such that combining data with different bin sizes is not ideal. This can be seen 
at the extremities of the spectra binned to different error bars by a mismatch of overlapping bins, although they agree within the errors.

ESPaDOnS was not designed for accurate continuum polarization measurements. Tests by CFHT staff have shown that for a given star, 
the level of continuum polarization can vary by as much as 10\%\footnote{http://www.cfht.hawaii.edu/Instruments/Spectroscopy/Espadons/ContiPolar/}\ 
and that furthermore, it is not flat as a function of wavelength. However, as the change is slow as a function of wavelength, the {\em difference} between 
the polarization in the lines and that in the neighbouring continuum is reliable. In Figures~1~and~2, the slow trend with wavelength which 
has been fitted by a straight line for $U/I$ and a low order polynomial for $Q/I$ (dashed lines), is certainly instrumental. 
Note however that the general level of polarization detected here ($Q\sim6.2\%\  \&\  U\sim-1.5\%$) is roughly compatible with the value published by Schmidt (1988), $Q\sim6.4\%\  \&\  U\sim-1.1\%$ ($P\sim6.5\%\  \&\  \theta\sim95^o$). As will be seen below, this very likely is mostly interstellar in origin.

In Figures~1 and 2, a clear signal can be seen for all main emission lines present in the spectrum of WR~1 in both linear polarization Stokes parameters. In a plot of the relative Stokes parameters as a function of wavelength, a signal from interstellar polarization would not have such a signature. Indeed, the light, regardless of its origin (line or continuum), will get polarized by the same fraction when it travels through the interstellar medium along the line of sight,
yielding a flat spectrum in relative polarization. However if the continuum light from the star is polarized, then the relative Stokes parameters would show dilution of this polarized continuum by unpolarized line emission, which is exactly what is observed here. As mentioned above, for ESPaDOnS data, the actual level of continuum polarization is not reliable and an additional curvature caused by the instrument is observed as a function of wavelength. However, the difference between the polarization in a line and that of nearby continuum remains trustworthy.  Note that the above-mentioned curvature is not to be confused with the well-known Serkowski law \citep{1975ApJ...196..261S} describing the behaviour of interstellar polarization with wavelength. A Serkowski function cannot be fitted here in view if instrumental effects.

In the context of this work and because of the nature of ESPaDOns data, the quantity of interest is the {\it difference} between the value of the polarization in the line and the value in the neighbouring continuum. The detected 
signal is too weak for this star to study the detailed shape of the polarization line profile. To quantify the difference in polarization between line centre and the neighbouring continuum, a gaussian profile was fitted to each of the lines in $Q/I$ and $U/I$. Table~1 lists the values of the core intensity of these gaussians for the main lines of WR~1. 

\begin{table}
\begin{center}
\caption{Linear Polarization Measurements for the Main Spectral Lines of WR~1 }
\vspace{0.5truecm}
\begin{tabular}{lccccc}
\tableline\tableline
Line & Q/I & U/I  &  $I/I_c$\\ 
& (\%) & (\%) & \\ 
\tableline
He{\sc ii}$\lambda$4542 & $-$0.145 & \ \ $+$0.180  & 1.9\\
N{\sc v}$\lambda$4603-19 & $-$0.181 & \ \ $+$0.216  & 2.4 \\
He{\sc ii}$\lambda$4686  & $-$0.315 &\ \ $+$0.358 & 9.9 \\
He{\sc ii}$\lambda$4861  & $-$0.132 & \ \ $+$0.305 & 2.0\\
N{\sc v}$\lambda$4945 & $-$0.092 & \ \ $+$0.126 & 1.25\\
He{\sc ii}$\lambda$5411  & $-$0.169 & \ \ $+$0.252 & 2.7\\
C{\sc iv}$\lambda$5804  & $-$0.116 & \ \ $+$0.115 & 2.1\\
He{\sc i}$\lambda$5876  & $-$0.062 & \ \ $+$0.086 & 1.3\\
He{\sc ii}$\lambda$6410  & $-$0.101 & \ \ $+$0.122 & 1.2\\
He{\sc ii}$\lambda$6560  & $-$0.192 & \ \ $+$0.226 & 3.6\\
He{\sc i}$\lambda$6678  & $-$0.070 & \ \ $+$0.086 & 1.4\\
N{\sc iv}$\lambda$7113  & $-$0.244 & \ \ $+$0.167 & 3.5\\
\tableline
\end{tabular}

\end{center}
\end{table}

Let the relative line strength, $R$ be defined as $R=(I_{line}+I_{cont})/I_{cont}$, where $I_{line}$ is the flux in the line only i.e. does not include continuum flux. Also, if the line polarization flux is assumed to be zero,  the polarization flux at the position of a spectral line, $Q_{line},U_{line}$ is simply equal to the polarization flux of the continuum, $Q_{cont},U_{cont}$. The difference between the relative polarization at the position of a spectral line and that if the neighbouring continuum is  then

$$q_{line}-q_{cont}={Q_{line}\over (I_{line}+I_{cont})}-{Q_{cont}\over I_{cont}}={Q_{line}\over R\,I_{cont}}-{Q_{cont}\over I_{cont}}={Q_{cont} \over I_{cont}} \left({1\over R}-1\right)=q_{cont}\left({1\over R}-1\right),$$

$$u_{line}-u_{cont}={U_{line}\over (I_{line}+I_{cont})}-{U_{cont}\over I_{cont}}={U_{line}\over R\,I_{cont}}-{U_{cont}\over I_{cont}}={U_{cont} \over I_{cont}}\left({1\over R}-1\right)=u_{cont}\left({1\over R}-1\right),$$ 

where $q$ and $u$ are the relative Stokes parameters.

These equations are exactly the same as those of \citet{1991ApJ...382..301S}. From them, it can be seen that the slope of a $q_{line}-q_{cont}$ ($u_{line}-u_{cont}$) versus $1/R$ plot gives $q_{cont}$ ($u_{cont}$) and that the intercept gives $-q_{cont}$ ($-u_{cont}$) . In Figure~3, we plot the difference between the relative Stokes parameters in the line and continuum versus the inverse of the relative line strength, $R$. Straight lines were fitted based on $\chi^2$ statistics. The values of the slopes and intercepts for both linear Stokes parameters with their associated uncertainties are indicated on the plots.

\begin{figure}[!Hhtb]
\includegraphics[scale=0.85]{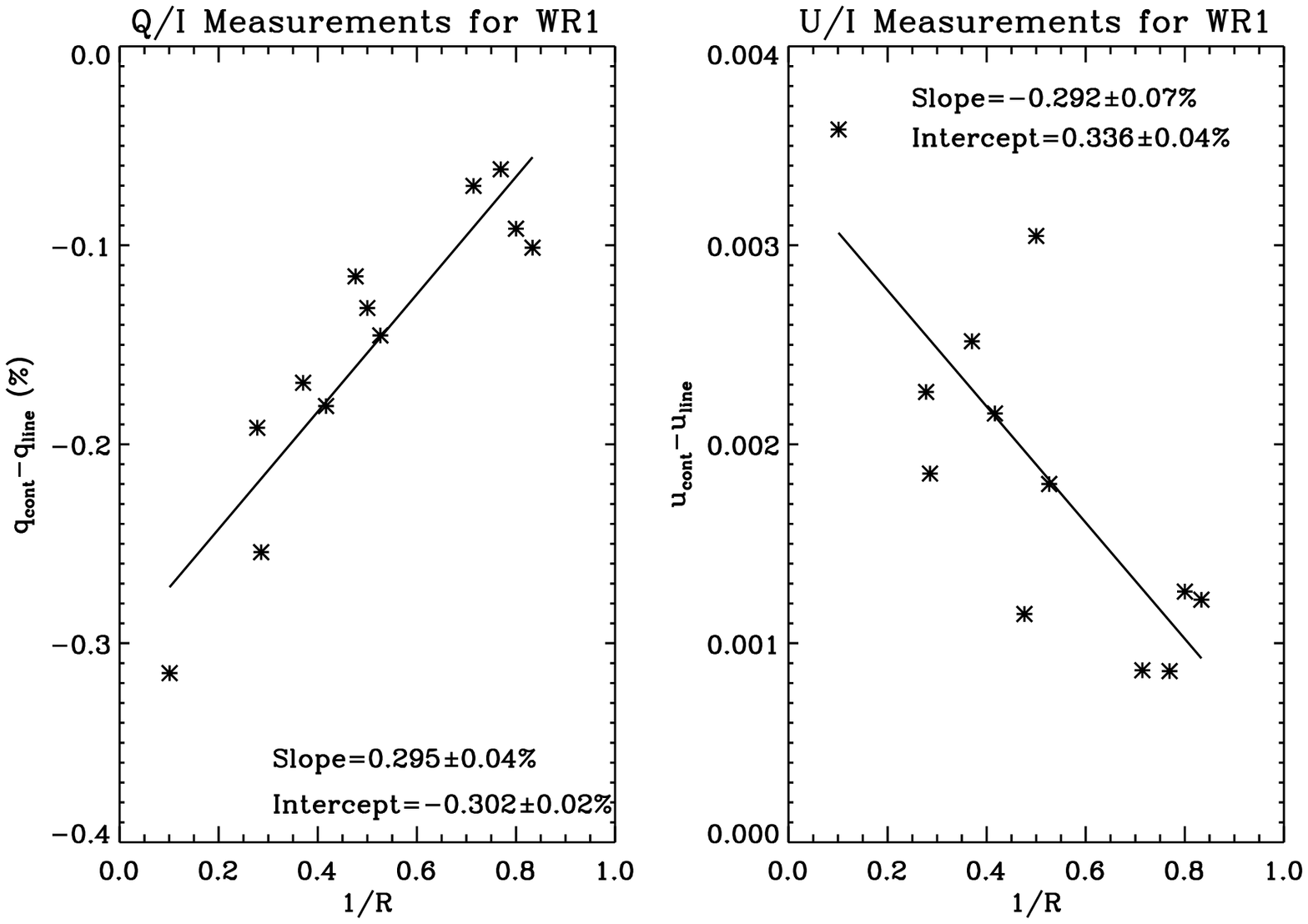}
\caption{ $q_{line}-q_{cont}$ and $u_{line}-u_{cont}$ versus 1/R plots for WR1. The slope and intercept of the straight line fitted to the data give the value of the continuum polarization of the star.}
\end{figure}

The final values of $q_{cont}$ and $u_{cont}$ are calculated by taking the mean, weighted by the inverse of the square of the uncertainty, of the slope and the intercept (multiplied by $-1$) of the above plots: $q_{cont}=0.301\pm 0.018\%$ and $u_{cont}=-0.325\pm 0.035\%$. This gives 
$$p_{cont}={P_{cont} \over I_{cont}}= \sqrt{{q_{cont}}^2 + {u_{cont}}^2}=0.443\pm 0.028\%$$
$$\theta={1\over 2}\, \tan^{-1}(u_{cont}/q_{cont})=-26.2\pm2.6^o$$

One can wonder what would happen if the assumption that the lines are completely unpolarized was incorrect. In such a case, we would have :
$$(I_{line}+I_{cont})q_{line}=I_{cont}q_{cont}+I_{line}q'_{line}$$
$$(I_{line}+I_{cont})u_{line}=I_{cont}u_{cont}+I_{line}u'_{line}$$
where the primed polarization values are those only in the line.

From this follows:
$$ q_{line}={{I_{cont}q_{cont}+ I_{line}q'_{line}}\over {I_{line}+I_{cont}}}$$ 
$$ u_{line}={{I_{cont}u_{cont}+ I_{line}u'_{line}}\over {I_{line}+I_{cont}}}$$ 

As $R=I_{cont}/(I_{line}+I_{cont})$, the $I_{line}=I_{cont}(R-1)$. We then have 

$$q_{line}-q_{cont}={q_{cont}\over R}+{{I_{cont}(R-1)} \over {I_{line}+I_{cont}}}q'_{line}-q_{cont}$$
$$u_{line}-u_{cont}={u_{cont}\over R}+{{I_{cont}(R-1)} \over {I_{line}+I_{cont}}}u'_{line}-u_{cont}$$

and finally

$$q_{line}-q_{cont}=(q_{cont}-q'_{line}){1\over R}-(q_{cont}-q'_{line})=(q_{cont}-q'_{line})\left({1\over R}-1\right)$$
$$u_{line}-u_{cont}=(u_{cont}-u'_{line}){1\over R}-(u_{cont}-u'_{line})=(u_{cont}-u'_{line})\left({1\over R}-1\right)$$

If $q'_{line}$ and $u'_{line}$ are constant for all lines, a plot of  $q_{line}-q_{cont}$ ($u_{line}-u_{cont}$) versus $1/R$ would still give a straight line. However, in view of the well-known stratification of WR stellar winds \citep[e.g.][]{2000ApJS..126..469H}, this is unlikely. The lines formed closer to the base of the wind should have higher polarizations in view of the higher electron densities in these regions. If the situation were that simple, this would lead to a systematic trend in the deviation of the points from the straight-line which assumes zero line polarization. However, it will most likely be more complicated since for axisymmetric winds, the mass-loss and wind velocity become latitude dependent. Also, the star could have a different brightness at the pole and equator. Finally, depending if the ion is a dominant one or not, it may become more or less prominent in the low or high density flow {\bf \citep[e.g.][]{1991A&A...244L...5L, 2000A&A...358..956P, 2006MNRAS.371..343I}. }Therefore, the effect of line polarization would likely be to randomly add scatter to the linear relation.

\section{Discussion}
The data presented in this paper represents a new detection of an asymmetric wind of a WR stars using spectropolarimetry. How does this compare with previous observations in terms of absolute level of polarization and type of WR star in which it appears and what can be concluded as to the origin of the asymmetry?

\citet{1988prco.book..641S} published spectropolarimetric observations of several WR stars obtained with the Image-Dissector Scanner spectropolarimeter at Lick Observatory in the early 80s. The original bin size was $\sim$10 \AA\ but the final polarization values were based on bins of 15-20 \AA\ wide. For WR1, no line depolarization was detected and the final overall value of $P$ given by the author is $P=6.24\% \pm0.03\%$ . However this error bar is most certainly the error on the mean taken over the entire wavelength range of their data and is not representative of the scatter in the spectrum.
Indeed, an eye estimation from a visual inspection of the $P$ spectrum plot yields (for lack of a better means) a scatter around the mean  (on the plot with 10 \AA\ bins, however) which is closer to $2\sigma \sim 1\%$ indicating that the error bar on individual bins is closer to 0.5\%\ instead of 0.03\%. This would be more compatible with the results presented in the present work. Indeed, a line depolarization of 0.443\%\  would have been easily detected if the error bar on each individual wavelength bin would have been 0.03\%. On the other hand, such a small value would be lost in the scatter in the plot provided in the paper.

\citet{1998MNRAS.296.1072H} presented a compilation of 29 WR stars with accurate (0.05\%) spectropolarimetry and found only 6 that showed line depolarization. With this new detection, there are now seven WR stars for which the line effect has been reported:  WR1 (WN4b), WR6 (WN6b), WR16 (WN8h), WR40 (WN8h), WR134 (WN6b), WR136 (WN6b(h)) and WR137 (WC7pd+O4-5). Of these stars, three, all broad-lined WN stars have been shown to present periodic variations thought to be caused by the presence of CIRs in their wind, WR1 \citep{2010ApJ...716..929C}, WR 6 \citep{1995ApJ...452L..57S, 1997ApJ...482..470M} and WR134 \citep{1994Ap&SS.221..155M, 1999ApJ...518..428M}. WR 136, the only other broad-line WN star, is a special case since line depolarization has been detected only once by \citet{1988BAAS...20.1013W}. \citet{1988prco.book..641S}, \citet{1994Ap&SS.221..347S} and \citet{1998MNRAS.296.1072H} also observed this star but made no detection of line depolarization. WR 137, a long-period WC7pd+O9 binary (P=4766 days) was found to also show photometric variability with a very short 0.83-day period \citep{2005MNRAS.360..141L}. Those authors suggest that the short term variability might  also be associated with the presence of a CIR, although that has not yet been demonstrated with spectroscopic observations. Finally, the two last stars, WR16 and WR40 are part of the WN8 sub-class which is well-known for showing high levels of variability in photometry, polarimetry and spectroscopy \citep[e.g.][]{1986AJ.....92..952M,1987ApJ...322..888D,1989ApJ...347.1034R,1995AJ....109..817A,1998MNRAS.294..642M}. 

The seven stars listed above therefore have winds that did not have spherical symmetry at the time the spectropolarimetric observations were obtained. However, the detected continuum polarization is not necessarily constant in time. Indeed, most WR stars are found to show variable continuum polarization. In a series of papers \citep{1987ApJ...322..870S,1987ApJ...322..888D,1988ApJ...330..286S,1989ApJ...343..426D,1989ApJ...347.1034R,1990ApJ...359..211R,1992ApJ...386..288D}, the broadband continuum polarization behaviour with time of a sample of WR stars was studied. A correlation between the degree of random, intrinsic scatter in polarization and both the spectral sub-type and wind terminal velocity was found. These two parameters are, however, found to be correlated \citep[e.g.][]{2007ARA&A..45..177C}. Faster-wind WR stars are found to show a lower level of intrinsic variability. The random polarization variability in this case is thought to be a consequence of inhomogeneities in the wind, which is considered to be ubiquitous for this type of outflow \citep[e.g.][]{2008AJ....136..548L}. At any given time the blob distribution generates a net polarization level which fluctuates randomly in time because the global blob configuration changes. Several theoretical studies have been carried out on this subject. Initial models by \citet{1996A&A...306..519R} \citep[but see also][]{1995A&A...295..725B} were unable to reproduce the observed ratio of polarimetric to photometric variability $R=\sigma_{pol}/\sigma_{phot}\simeq0.05$ without invoking very high densities within the blobs. {\it 3-D} simulations of random (in time and distribution) blob ejection by \citet{2000A&A...357..233L} and \citet{2009RAA.....9..558L} including a $\beta$-law for the wind velocity distribution and stellar occultation effects where not only able to reproduce the observed value of $R$ without a very high value of the blob density but also to carry out estimates of quantities such as the  $\beta$-law exponent, the number of blobs and the total mass-loss rates in blobs. A similar model was applied to LBVs by \citet{2007A&A...469.1045D}. Recent Monte-Carlo simulations by \citet{2012AIPC.1429..278T} using a novel semi-analytic method concludes that the single scattering approximation made in previous models overestimates the mean polarization of optically thick winds such as those of WR stars, even if the clumps themselves are optically thin.

There is no reason to believe that random, intrinsic scatter in polarization cannot happen in a wind that is asymmetric. Therefore, the scatter could be centred on zero linear polarization if the wind is spherically symmetric or on another value if it is not. Of course, one must remove the interstellar polarization beforehand. One might think that with one snapshot observation in spectropolarimetry, it is impossible to conclude if the continuum polarization we are measuring is from a wind asymmetry that is constant in time or if it is due to a random arrangement of blobs in the wind that is generating a residual polarization. However, the level of continuum linear polarization ($\sim$ 0.5\%) measured for the three broad-line WN stars shown to have CIRs in their wind (WR1, WR6 and WR134) is most likely due to a large-scale structure. Indeed, the levels of scatter generated from random fluctuations ranges from $\sim$0.01-0.04\%  for early type WN stars to $\sim$0.08-0.15\% for WN8 stars. This is much smaller than the measured values in those 3 cases.  Therefore it is extremely unlikely that the residual polarization measured for these stars stems from a lucky configuration of blobs in the wind. More likely, it is the sign of a large-scale structure in the wind. Such levels of variability can be produced by either a flattened wind \citep{1998MNRAS.296.1072H} or a CIR in the wind \citep{2009AJ....137.3339I}

WR136 might be a similar case. \citet{1988BAAS...20.1013W} measured the linear polarization of this star in 10 broad filters from $\sim$ 4200 to 8500 \AA. They found a dip of $\Delta P\sim$0.5\%\  in one of them centred at 5900 \AA\ (100\AA\ wide). This filter includes the He{\sc i}$\lambda$5876 line and therefore they concluded that the star showed line depolarization.
However, this was not observed by 3 other authors in independent observations (see above). One possible explanation is that the line-depolarization for this star or in other words the polarization of its continuum light is a transient phenomenon. If the linear polarization is due to the presence of a CIR for example, it might mean that they occur only occasionally in this star. One supporting observation for CIRs in the wind of this star is the detection of enhanced P~Cygni  absorption at velocities in excess of the terminal velocity of the wind in the C{\sc iv}$\lambda$1550, He{\sc ii}$\lambda$1640 and N{\sc iv}$\lambda$1718 ultraviolet (UV) P~Cygni  profiles of this star by \citet{1989A&A...226..249S}. The variations were observed over a period of 24 hours in 1987. Two other datasets, one in 1981 and the other in 1982, covering a similar period of time, were found not to show enhanced P~Cygni absorption and indeed no variations at all. This type of blue-edge variability is thought to be associated with the presence of CIRs in the wind \citep{1996ApJ...462..469C}.

The WC7pd+ O9 binary WR137 shows a modest line linear depolarization of 0.1-0.2\%\ \citep[][de la Chevroti\`ere et al. submitted]{1998MNRAS.296.1072H}. WR+O binaries are well known to show significant levels of linear polarization variability in phase with the orbital period \citep{1987ApJ...322..870S} as the O-star continuum acts as an asymmetric light source for the free electron distribution in the WR wind. However, in this case, the 13.05 yr orbit \citep{2001MNRAS.324..156W} is much too large for such an effect to take place. Therefore as already mentioned by \citet{1998MNRAS.296.1072H} the continuum polarization is more likely intrinsic to the WR wind. The random broadband polarization variability reported by \citet{1989ApJ...347.1034R} for this star is of a similar level. Therefore, it is possible that the continuum polarization of the WR star in this system is simply due to blobs in the wind. However, one must also consider that because this is a WR+O binary, the relative polarization measurements of the line depolarization are diluted because although the polarization is likely to occur in the wind of one star, we are dividing that polarized flux by the total light of the two stars, thereby diluting the line-depolarization. The original linear polarization in the wind is therefore most certainly higher and might be caused by a global asymmetry in the wind. This result is possibly supported by the work of \citet{2005MNRAS.360..141L} who found a very short 0.83d period in small-amplitude variations of the absorption throughs of the C{\sc iv}$\lambda\lambda$5802/5812 and He{\sc i}$\lambda$5876 P~Cygni profiles, which they associate with CIRs or non-radial pulsations. If it were CIRs in the WR wind, it could explain the continuum polarization level observed by \citet{1998MNRAS.296.1072H}.
 
Finally, the line depolarization measured for the two WN8 stars WR16 and WR40 \citep{2011A&A...536L..10V} on the other hand is very small, of the order of $\sim$0.1\%. This could very well be a random polarization level due to blobs, particularly because the scatter in polarization is very high for these WN8 type stars.

\section{Conclusion}
In summary, of the seven stars showing line depolarization and therefore having some level of continuum polarization, three (WR1, WR6 \&\ WR134) and possibly four (including WR136) have quite a large amplitude of $\sim$0.5\%\ that can be explained either by a flattened wind or the presence of large-scale structure such as CIRs. For WR1, WR6 and WR134, the presence of CIRs in the wind is supported by kinematic evidence from periodic variations in optical emission lines. Therefore this hypothesis is preferred. For WR136, the phenomenon seems to be more elusive and needs to be confirmed. These four stars are all broad-line early WN stars. Based on the work of \citet{1993A&A...274..397H}, \citet{1996MNRAS.281..163S} introduced the {\it b} subscript (together with others) as an extension of the classical WR star classification system to distinguish those that have a He{\sc ii}$\lambda$5411 equivalent width greater than 40 \AA. The usefulness of this was questioned by \citet{1999NewA....4..489C} because it was found that the observed distribution of line width did not have a sufficiently clear delimitation at the 40 \AA\ level \citep[but see][] {2000NewA....5..423S}. So it is not clear if these stars have a peculiar characteristic that can help explain the presence of CIRs in their wind or if the presence of CIRs in these three stars with a WNb classification is just a simple coincidence. One difference is that stars with the WNb classification are never found to have hydrogen while those that are simply classified WN sometimes are found to have hydrogen and sometimes not. A final point to consider is the recent strong correlation found by \citet{2011A&A...536L..10V} and \citet{2012A&A...547A..83G} between line depolarization in WR stars and the presence of an ejected nebula. As ejected nebula are predicted to last only for $\sim 10^5$ years \citep{2005A&A...444..837V}, these authors associated line depolarization with youth in the WR phase. This might indicate that CIRs only occur in WR winds when they have just recently reached this evolutionary phase.

Stars that have CIRs in their winds are extremely useful because if the structures are pinned to their surfaces, the periods of the spectroscopic variations they generate together
with an estimate of the stars' radii allows us to measure the rotation velocities of the stars, which cannot be obtained by regular means as there are no photospheric lines in the spectrum. However, CIRs are structures that appear and dissipate on an unknown timescale which leads to epoch-dependant
variations. Therefore, they are extremely difficult to identify. If one can establish a link between line depolarization of WR stars and the presence of a CIR, this can provide an easier means of identifying them. Although detailed spectroscopic follow-up observations would still be required to determine the rotation period, this would certainly represent a significant improvement on the present situation.

\acknowledgments

First, NSL would like to thank the referee, Rico Ignace for a thorough evaluation of the manuscript, including many constructive comments that led to the improvement of this paper. NSL also thanks Nadine Manset from the CFHT for continuous support in data reduction issues and Anthony F.J. Moffat and Noel Richardson for comments on the manuscript. NSL also acknowledges financial support from the Natural Science and Engineering Research Council (NSERC) of Canada.

{\it Facilities:} \facility{CFHT}

\bibliography{stlouis}

\begin{thebibliography}{}
\expandafter\ifx\csname natexlab\endcsname\relax\def\natexlab#1{#1}\fi

\bibitem[{{Antokhin} {et~al.}(1995){Antokhin}, {Bertrand}, {Lamontagne},
  {Moffat}, \& {Matthews}}]{1995AJ....109..817A}
{Antokhin}, I., {Bertrand}, J.-F., {Lamontagne}, R., {Moffat}, A.~F.~J., \&
  {Matthews}, J. 1995, \aj, 109, 817

\bibitem[{{Bouret} {et~al.}(2005){Bouret}, {Lanz}, \&
  {Hillier}}]{2005A&A...438..301B}
{Bouret}, J.-C., {Lanz}, T., \& {Hillier}, D.~J. 2005, \aap, 438, 301

\bibitem[{{Brott} {et~al.}(2011){Brott}, {de Mink}, {Cantiello}, {Langer}, {de
  Koter}, {Evans}, {Hunter}, {Trundle}, \& {Vink}}]{2011A&A...530A.115B}
{Brott}, I., {de Mink}, S.~E., {Cantiello}, M., {et~al.} 2011, \aap, 530, A115

\bibitem[{{Brown} {et~al.}(1995){Brown}, {Richardson}, {Antokhin}, {Robert},
  {Moffat}, \& {St-Louis}}]{1995A&A...295..725B}
{Brown}, J.~C., {Richardson}, L.~L., {Antokhin}, I., {et~al.} 1995, \aap, 295,
  725

\bibitem[{{Chen{\'e}} \& {St-Louis}(2010)}]{2010ApJ...716..929C}
{Chen{\'e}}, A.-N., \& {St-Louis}, N. 2010, \apj, 716, 929

\bibitem[{{Chiosi} \& {Maeder}(1986)}]{1986ARA&A..24..329C}
{Chiosi}, C., \& {Maeder}, A. 1986, \araa, 24, 329

\bibitem[{{Conti}(1999)}]{1999NewA....4..489C}
{Conti}, P.~S. 1999, \na, 4, 489

\bibitem[{{Cranmer} \& {Owocki}(1995)}]{1995ApJ...440..308C}
{Cranmer}, S.~R., \& {Owocki}, S.~P. 1995, \apj, 440, 308

\bibitem[{{Cranmer} \& {Owocki}(1996)}]{1996ApJ...462..469C}
---. 1996, \apj, 462, 469

\bibitem[{{Crowther}(2007)}]{2007ARA&A..45..177C}
{Crowther}, P.~A. 2007, \araa, 45, 177

\bibitem[{{Davies} {et~al.}(2007){Davies}, {Vink}, \&
  {Oudmaijer}}]{2007A&A...469.1045D}
{Davies}, B., {Vink}, J.~S., \& {Oudmaijer}, R.~D. 2007, \aap, 469, 1045

\bibitem[{{de la Chevroti{\`e}re} {et~al.}(2013){de la Chevroti{\`e}re},
  {St-Louis}, {Moffat}, \& {the MiMeS Collaboration}}]{2013ApJ...764..171D}
{de la Chevroti{\`e}re}, A., {St-Louis}, N., {Moffat}, A.~F.~J., \& {the MiMeS
  Collaboration}. 2013, \apj, 764, 171

\bibitem[{{Dessart} \& {Chesneau}(2002)}]{2002A&A...395..209D}
{Dessart}, L., \& {Chesneau}, O. 2002, \aap, 395, 209

\bibitem[{{Donati} {et~al.}(1997){Donati}, {Semel}, {Carter}, {Rees}, \&
  {Collier Cameron}}]{1997MNRAS.291..658D}
{Donati}, J.-F., {Semel}, M., {Carter}, B.~D., {Rees}, D.~E., \& {Collier
  Cameron}, A. 1997, \mnras, 291, 658

\bibitem[{{Drissen} {et~al.}(1989){Drissen}, {Robert}, {Lamontagne}, {Moffat},
  {St-Louis}, {van Weeren}, \& {van Genderen}}]{1989ApJ...343..426D}
{Drissen}, L., {Robert}, C., {Lamontagne}, R., {et~al.} 1989, \apj, 343, 426

\bibitem[{{Drissen} {et~al.}(1992){Drissen}, {Robert}, \&
  {Moffat}}]{1992ApJ...386..288D}
{Drissen}, L., {Robert}, C., \& {Moffat}, A.~F.~J. 1992, \apj, 386, 288

\bibitem[{{Drissen} {et~al.}(1987){Drissen}, {St.-Louis}, {Moffat}, \&
  {Bastien}}]{1987ApJ...322..888D}
{Drissen}, L., {St.-Louis}, N., {Moffat}, A.~F.~J., \& {Bastien}, P. 1987,
  \apj, 322, 888

\bibitem[{{Gayley} \& {Ignace}(2010)}]{2010ApJ...708..615G}
{Gayley}, K.~G., \& {Ignace}, R. 2010, \apj, 708, 615

\bibitem[{{Gr{\"a}fener} {et~al.}(2012){Gr{\"a}fener}, {Vink}, {Harries}, \&
  {Langer}}]{2012A&A...547A..83G}
{Gr{\"a}fener}, G., {Vink}, J.~S., {Harries}, T.~J., \& {Langer}, N. 2012,
  \aap, 547, A83

\bibitem[{{Hamann} \& {Koesterke}(1998)}]{1998A&A...335.1003H}
{Hamann}, W.-R., \& {Koesterke}, L. 1998, \aap, 335, 1003

\bibitem[{{Hamann} {et~al.}(1993){Hamann}, {Koesterke}, \&
  {Wessolowski}}]{1993A&A...274..397H}
{Hamann}, W.~R., {Koesterke}, L., \& {Wessolowski}, U. 1993, \aap, 274, 397

\bibitem[{{Harries} {et~al.}(1998){Harries}, {Hillier}, \&
  {Howarth}}]{1998MNRAS.296.1072H}
{Harries}, T.~J., {Hillier}, D.~J., \& {Howarth}, I.~D. 1998, \mnras, 296, 1072

\bibitem[{{Herald} {et~al.}(2000){Herald}, {Schulte-Ladbeck}, {Eenens}, \&
  {Morris}}]{2000ApJS..126..469H}
{Herald}, J.~E., {Schulte-Ladbeck}, R.~E., {Eenens}, P.~R.~J., \& {Morris}, P.
  2000, \apjs, 126, 469

\bibitem[{{Howarth} \& {Prinja}(1989)}]{1989ApJS...69..527H}
{Howarth}, I.~D., \& {Prinja}, R.~K. 1989, \apjs, 69, 527

\bibitem[{{Ignace} \& {Brimeyer}(2006)}]{2006MNRAS.371..343I}
{Ignace}, R., \& {Brimeyer}, A. 2006, \mnras, 371, 343

\bibitem[{{Ignace} {et~al.}(2009){Ignace}, {Hubrig}, \&
  {Sch{\"o}ller}}]{2009AJ....137.3339I}
{Ignace}, R., {Hubrig}, S., \& {Sch{\"o}ller}, M. 2009, \aj, 137, 3339

\bibitem[{{Lamers} \& {Pauldrach}(1991)}]{1991A&A...244L...5L}
{Lamers}, H.~J.~G., \& {Pauldrach}, A.~W.~A. 1991, \aap, 244, L5

\bibitem[{{Langer}(2012)}]{2012ARA&A..50..107L}
{Langer}, N. 2012, \araa, 50, 107

\bibitem[{{Lef{\`e}vre} {et~al.}(2005){Lef{\`e}vre}, {Marchenko}, {L{\'e}pine},
  {Moffat}, {Acker}, {Harries}, {Annuk}, {Bohlender}, {Demers}, {Grosdidier},
  {Hill}, {Morrison}, {Knauth}, {Skalkowski}, \& {Viti}}]{2005MNRAS.360..141L}
{Lef{\`e}vre}, L., {Marchenko}, S.~V., {L{\'e}pine}, S., {et~al.} 2005, \mnras,
  360, 141

\bibitem[{{L{\'e}pine} \& {Moffat}(1999)}]{1999ApJ...514..909L}
{L{\'e}pine}, S., \& {Moffat}, A.~F.~J. 1999, \apj, 514, 909

\bibitem[{{L{\'e}pine} \& {Moffat}(2008)}]{2008AJ....136..548L}
---. 2008, \aj, 136, 548

\bibitem[{{Li} {et~al.}(2000){Li}, {Brown}, {Ignace}, {Cassinelli}, \&
  {Oskinova}}]{2000A&A...357..233L}
{Li}, Q., {Brown}, J.~C., {Ignace}, R., {Cassinelli}, J.~P., \& {Oskinova},
  L.~M. 2000, \aap, 357, 233

\bibitem[{{Li} {et~al.}(2009){Li}, {Cassinelli}, {Brown}, \&
  {Ignace}}]{2009RAA.....9..558L}
{Li}, Q.-K., {Cassinelli}, J.~P., {Brown}, J.~C., \& {Ignace}, R. 2009,
  Research in Astronomy and Astrophysics, 9, 558

\bibitem[{{Maeder}(1981)}]{1981A&A...102..401M}
{Maeder}, A. 1981, \aap, 102, 401

\bibitem[{{Maeder}(1999)}]{1999A&A...347..185M}
---. 1999, \aap, 347, 185

\bibitem[{{Maeder} \& {Meynet}(2000)}]{2000ARA&A..38..143M}
{Maeder}, A., \& {Meynet}, G. 2000, \araa, 38, 143

\bibitem[{{Maeder} \& {Meynet}(2004)}]{2004A&A...422..225M}
---. 2004, \aap, 422, 225

\bibitem[{{Marchenko} {et~al.}(1998){Marchenko}, {Moffat}, {Eversberg},
  {Morel}, {Hill}, {Tovmassian}, \& {Seggewiss}}]{1998MNRAS.294..642M}
{Marchenko}, S.~V., {Moffat}, A.~F.~J., {Eversberg}, T., {et~al.} 1998, \mnras,
  294, 642

\bibitem[{{McCandliss} {et~al.}(1994){McCandliss}, {Bohannan}, {Robert}, \&
  {Moffat}}]{1994Ap&SS.221..155M}
{McCandliss}, S.~R., {Bohannan}, B., {Robert}, C., \& {Moffat}, A.~F.~J. 1994,
  \apss, 221, 155

\bibitem[{{Meynet} {et~al.}(2011){Meynet}, {Eggenberger}, \&
  {Maeder}}]{2011A&A...525L..11M}
{Meynet}, G., {Eggenberger}, P., \& {Maeder}, A. 2011, \aap, 525, L11

\bibitem[{{Meynet} \& {Maeder}(2000)}]{2000A&A...361..101M}
{Meynet}, G., \& {Maeder}, A. 2000, \aap, 361, 101

\bibitem[{{Meynet} {et~al.}(1994){Meynet}, {Maeder}, {Schaller}, {Schaerer}, \&
  {Charbonnel}}]{1994A&AS..103...97M}
{Meynet}, G., {Maeder}, A., {Schaller}, G., {Schaerer}, D., \& {Charbonnel}, C.
  1994, \aaps, 103, 97

\bibitem[{{Moffat}(1982)}]{1982IAUS...99..263M}
{Moffat}, A.~F.~J. 1982, in IAU Symposium, Vol.~99, Wolf-Rayet Stars:
  Observations, Physics, Evolution, ed. C.~W.~H. {De Loore} \& A.~J. {Willis},
  263--273

\bibitem[{{Moffat}(1983)}]{1983wrsp.conf...13M}
{Moffat}, A.~F.~J. 1983, in Wolf-Rayet stars: Progenitors of supernovae?, ed.
  M.-C. {Lortet} \& A.~{Pitault}, 13

\bibitem[{{Moffat} {et~al.}(1988){Moffat}, {Drissen}, {Lamontagne}, \&
  {Robert}}]{1988ApJ...334.1038M}
{Moffat}, A.~F.~J., {Drissen}, L., {Lamontagne}, R., \& {Robert}, C. 1988,
  \apj, 334, 1038

\bibitem[{{Moffat} \& {Shara}(1986)}]{1986AJ.....92..952M}
{Moffat}, A.~F.~J., \& {Shara}, M.~M. 1986, \aj, 92, 952

\bibitem[{{Morel} {et~al.}(1997){Morel}, {St-Louis}, \&
  {Marchenko}}]{1997ApJ...482..470M}
{Morel}, T., {St-Louis}, N., \& {Marchenko}, S.~V. 1997, \apj, 482, 470

\bibitem[{{Morel} {et~al.}(1999){Morel}, {Marchenko}, {Eenens}, {Moffat},
  {Koenigsberger}, {Antokhin}, {Eversberg}, {Tovmassian}, {Hill}, {Cardona}, \&
  {St-Louis}}]{1999ApJ...518..428M}
{Morel}, T., {Marchenko}, S.~V., {Eenens}, P.~R.~J., {et~al.} 1999, \apj, 518,
  428

\bibitem[{{Mullan}(1986)}]{1986A&A...165..157M}
{Mullan}, D.~J. 1986, \aap, 165, 157

\bibitem[{{Nugis} {et~al.}(1998){Nugis}, {Crowther}, \&
  {Willis}}]{1998A&A...333..956N}
{Nugis}, T., {Crowther}, P.~A., \& {Willis}, A.~J. 1998, \aap, 333, 956

\bibitem[{{Petrenz} \& {Puls}(2000)}]{2000A&A...358..956P}
{Petrenz}, P., \& {Puls}, J. 2000, \aap, 358, 956

\bibitem[{{Richardson} {et~al.}(1996){Richardson}, {Brown}, \&
  {Simmons}}]{1996A&A...306..519R}
{Richardson}, L.~L., {Brown}, J.~C., \& {Simmons}, J.~F.~L. 1996, \aap, 306,
  519

\bibitem[{{Robert} {et~al.}(1989){Robert}, {Moffat}, {Bastien}, {Drissen}, \&
  {St.-Louis}}]{1989ApJ...347.1034R}
{Robert}, C., {Moffat}, A.~F.~J., {Bastien}, P., {Drissen}, L., \& {St.-Louis},
  N. 1989, \apj, 347, 1034

\bibitem[{{Robert} {et~al.}(1990){Robert}, {Moffat}, {Bastien}, {St.-Louis}, \&
  {Drissen}}]{1990ApJ...359..211R}
{Robert}, C., {Moffat}, A.~F.~J., {Bastien}, P., {St.-Louis}, N., \& {Drissen},
  L. 1990, \apj, 359, 211

\bibitem[{{Schmidt}(1988)}]{1988prco.book..641S}
{Schmidt}, G.~D. 1988, {Spectropolarimetry as a probe of the structure of
  Wolf-Rayet envelopes}, ed. G.~V. {Coyne}, A.~M. {Magalhaes}, A.~F. {Moffat},
  R.~E. {Schulte-Ladbeck}, \& S.~{Tapia}, 641--654

\bibitem[{{Schulte-Ladbeck}(1994)}]{1994Ap&SS.221..347S}
{Schulte-Ladbeck}, R.~E. 1994, \apss, 221, 347

\bibitem[{{Schulte-Ladbeck} {et~al.}(1991){Schulte-Ladbeck}, {Nordsieck},
  {Taylor}, {Nook}, {Bjorkman}, {Magalhaes}, \&
  {Anderson}}]{1991ApJ...382..301S}
{Schulte-Ladbeck}, R.~E., {Nordsieck}, K.~H., {Taylor}, M., {et~al.} 1991,
  \apj, 382, 301

\bibitem[{{Semel} \& {Li}(1996)}]{1996SoPh..164..417S}
{Semel}, M., \& {Li}, J. 1996, \solphys, 164, 417

\bibitem[{{Serkowski} {et~al.}(1975){Serkowski}, {Mathewson}, \&
  {Ford}}]{1975ApJ...196..261S}
{Serkowski}, K., {Mathewson}, D.~S., \& {Ford}, V.~L. 1975, \apj, 196, 261

\bibitem[{{Smith} {et~al.}(2000){Smith}, {Shara}, \&
  {Moffat}}]{2000NewA....5..423S}
{Smith}, L.~F., {Shara}, M., \& {Moffat}, A. 2000, \na, 5, 423

\bibitem[{{Smith} {et~al.}(1996){Smith}, {Shara}, \&
  {Moffat}}]{1996MNRAS.281..163S}
{Smith}, L.~F., {Shara}, M.~M., \& {Moffat}, A.~F.~J. 1996, \mnras, 281, 163

\bibitem[{{St-Louis} {et~al.}(2009){St-Louis}, {Chen{\'e}}, {Schnurr}, \&
  {Nicol}}]{2009ApJ...698.1951S}
{St-Louis}, N., {Chen{\'e}}, A.-N., {Schnurr}, O., \& {Nicol}, M.-H. 2009,
  \apj, 698, 1951

\bibitem[{{St-Louis} {et~al.}(1995){St-Louis}, {Dalton}, {Marchenko}, {Moffat},
  \& {Willis}}]{1995ApJ...452L..57S}
{St-Louis}, N., {Dalton}, M.~J., {Marchenko}, S.~V., {Moffat}, A.~F.~J., \&
  {Willis}, A.~J. 1995, \apjl, 452, L57

\bibitem[{{St.-Louis} {et~al.}(1987){St.-Louis}, {Drissen}, {Moffat},
  {Bastien}, \& {Tapia}}]{1987ApJ...322..870S}
{St.-Louis}, N., {Drissen}, L., {Moffat}, A.~F.~J., {Bastien}, P., \& {Tapia},
  S. 1987, \apj, 322, 870

\bibitem[{{St.-Louis} {et~al.}(1988){St.-Louis}, {Moffat}, {Drissen},
  {Bastien}, \& {Robert}}]{1988ApJ...330..286S}
{St.-Louis}, N., {Moffat}, A.~F.~J., {Drissen}, L., {Bastien}, P., \& {Robert},
  C. 1988, \apj, 330, 286

\bibitem[{{St-Louis} {et~al.}(1989){St-Louis}, {Smith}, {Stevens}, {Willis},
  {Garmany}, \& {Conti}}]{1989A&A...226..249S}
{St-Louis}, N., {Smith}, L.~J., {Stevens}, I.~R., {et~al.} 1989, \aap, 226, 249

\bibitem[{{Townsend}(2012)}]{2012AIPC.1429..278T}
{Townsend}, R. 2012, in American Institute of Physics Conference Series, Vol.
  1429, American Institute of Physics Conference Series, ed. J.~L. {Hoffman},
  J.~{Bjorkman}, \& B.~{Whitney}, 278--281

\bibitem[{{van Marle} {et~al.}(2005){van Marle}, {Langer}, \&
  {Garc{\'{\i}}a-Segura}}]{2005A&A...444..837V}
{van Marle}, A.~J., {Langer}, N., \& {Garc{\'{\i}}a-Segura}, G. 2005, \aap,
  444, 837

\bibitem[{{Vink} {et~al.}(2011){Vink}, {Gr{\"a}fener}, \&
  {Harries}}]{2011A&A...536L..10V}
{Vink}, J.~S., {Gr{\"a}fener}, G., \& {Harries}, T.~J. 2011, \aap, 536, L10

\bibitem[{{Vreux}(1985)}]{1985PASP...97..274V}
{Vreux}, J.-M. 1985, \pasp, 97, 274

\bibitem[{{Whitney} {et~al.}(1988){Whitney}, {Schulte-Ladbeck}, {Aspin},
  {Meade}, {Anderson}, {Clayton}, {Murison}, {Nordsieck}, {Nook}, \&
  {Schutt}}]{1988BAAS...20.1013W}
{Whitney}, B.~A., {Schulte-Ladbeck}, R.~E., {Aspin}, C., {et~al.} 1988, in
  Bulletin of the American Astronomical Society, Vol.~20, Bulletin of the
  American Astronomical Society, 1013

\bibitem[{{Williams} {et~al.}(2001){Williams}, {Kidger}, {van der Hucht},
  {Morris}, {Tapia}, {Perinotto}, {Morbidelli}, {Fitzsimmons}, {Anthony},
  {Caldwell}, {Alonso}, \& {Wild}}]{2001MNRAS.324..156W}
{Williams}, P.~M., {Kidger}, M.~R., {van der Hucht}, K.~A., {et~al.} 2001,
  \mnras, 324, 156

\bibitem[{{Woosley} {et~al.}(1993){Woosley}, {Langer}, \&
  {Weaver}}]{1993ApJ...411..823W}
{Woosley}, S.~E., {Langer}, N., \& {Weaver}, T.~A. 1993, \apj, 411, 823

\bibitem[{{Yoon} {et~al.}(2012){Yoon}, {Dierks}, \&
  {Langer}}]{2012A&A...542A.113Y}
{Yoon}, S.-C., {Dierks}, A., \& {Langer}, N. 2012, \aap, 542, A113

\end{thebibliography}
\clearpage

\end{document}